\documentclass{webofc}
\usepackage[varg]{txfonts}

\begin{document}

\title{Overview of the distributed image processing infrastructure to produce the Legacy Survey of Space and Time}

\author{\firstname{Fabio}    \lastname{Hernandez}\inst{1}\fnsep\thanks{\email{fabio@in2p3.fr}} \and
        \firstname{George}   \lastname{Beckett}\inst{2} \and
        \firstname{Peter}    \lastname{Clark}\inst{2} \and
        \firstname{Matt}     \lastname{Doidge}\inst{4} \and
        \firstname{Tim}      \lastname{Jenness}\inst{5} \and
        \firstname{Edward}   \lastname{Karavakis}\inst{6} \and
        \firstname{Quentin}  \lastname{Le Boulc'h}\inst{1} \and
        \firstname{Peter}    \lastname{Love}\inst{4} \and
        \firstname{Gabriele} \lastname{Mainetti}\inst{1} \and
        \firstname{Timothy}  \lastname{Noble}\inst{7} \and
        \firstname{Brandon}  \lastname{White}\inst{3} \and
        \firstname{Wei}      \lastname{Yang}\inst{8}
}

\institute{
    CNRS, CC-IN2P3, 21 avenue Pierre de Coubertin, CS70202, 69627 Villeurbanne CEDEX, France \and
    University of Edinburgh, Edinburgh, UK \and
    Fermi National Accelerator Laboratory, Batavia, IL, USA \and
    Lancaster University, Lancaster, UK \and
    Vera C. Rubin Observatory Project Office, 950 N. Cherry Ave., Tucson, AZ 85719, USA \and
    Brookhaven National Laboratory, Upton, NY, USA \and
    STFC, Harwell, UK \and
    SLAC National Accelerator Laboratory, 2575 Sand Hill Rd., Menlo Park, CA 94025, USA
}

\abstract{%
The Vera C.\ Rubin Observatory is preparing to execute the most ambitious astronomical survey ever attempted, the Legacy Survey of Space and Time (LSST). Currently the final phase of construction is under way in the Chilean Andes, with the Observatory's ten-year science mission scheduled to begin in 2025. Rubin's 8.4-meter telescope will nightly scan the southern hemisphere collecting imagery in the wavelength range 320–1050 nm covering the entire observable sky every 4 nights using a 3.2~gigapixel camera, the largest imaging device ever built for astronomy. Automated detection and classification of celestial objects will be performed by sophisticated algorithms on high-resolution images to progressively produce an astronomical catalog eventually composed of 20 billion galaxies and 17 billion stars and their associated physical properties.

In this article we present an overview of the system currently being constructed to perform data distribution as well as the annual campaigns which reprocess the entire image dataset collected since the beginning of the survey. These processing campaigns will utilize computing and storage resources provided by three Rubin data facilities (one in the US and two in Europe). Each year a Data Release will be produced and disseminated to science collaborations for use in studies comprising four main science pillars: probing dark matter and dark energy, taking inventory of solar system objects, exploring the transient optical sky and mapping the Milky Way.

Also presented is the method by which we leverage some of the common tools and best practices used for management of large-scale distributed data processing projects in the high energy physics and astronomy communities. We also demonstrate how these tools and practices are utilized within the Rubin project in order to overcome the specific challenges faced by the Observatory.
}

\maketitle

\section{Introduction}
\label{introduction}
Located in the Chilean Andes and currently in its final phase of construction, the Vera C.\ Rubin Observatory is preparing to execute the Legacy Survey of Space and Time (LSST), the most ambitious astronomical survey ever attempted. Over the course of the Observatory's ten-year science mission, its 8.4-meter telescope will scan the southern  hemisphere collecting imagery in the wavelength range 320–1050 nm covering the entire observable sky every 4 nights using a 3.2\,gigapixel camera. Sophisticated algorithms applied to high-resolution images will perform automated detection and classification of celestial objects, progressively populating numerous physical parameters comprising an astronomical catalog of 20 billion galaxies and 17 billion stars. Science-ready images and the astronomical catalog will be regularly delivered by the Observatory to science collaborations for their studies in the four science pillars: probing dark energy and dark matter, taking an inventory of the solar system, exploring the transient optical sky and mapping the Milky Way\,\cite{Ivezic:2019}.

This paper is structured as follows. In section \ref{section-data-products} we present a high-level view of the data products that will be delivered by the Observatory. Section \ref{section-image-processing} covers the tools being developed for processing the images and section \ref{section-processing-infrastructure} the data processing infrastructure the project is deploying to generate those products.

\section{Data products}
\label{section-data-products}

The Rubin Observatory will deliver several kinds of data products as well as services for archiving and dissemination of those products to LSST Science Collaborations\,\cite{melissa_graham_2022_7011229}.

Raw images collected each observing night are processed within 60 seconds of their capture in order to generate and emit alerts for transient detection, in a process known as Prompt Processing. In addition, on an annual cadence Data Release Processing will entail a reprocessing of the entirety of the Rubin raw image set recorded since the beginning of the survey, producing a new Data Release for dissemination to the scientific community. Each Data Release includes, in addition to raw and calibration images, science-ready images which have been reprocessed with updated scientific algorithms, as well as a catalog with the properties of the astrophysical objects detected on the input images.

Five petabytes of new raw images will be recorded each year. The volume of released data products generated by the annual processing of the accumulated set of raw images is on average 2.3 times the size of the input data set for that year and is estimated to reach more than one hundred petabytes by the end of the survey.\footnote{This figure does not include the size of the intermediate data products generated for the needs of the processing but not included in the annual data release.} Released data products include processed visit images, coadd images and tabular data used to populate the astronomical catalog database\,\cite{LSE-163}. Over the ten year-long survey the volume of data released for science analysis is estimated to increase by one order of magnitude\,\cite{DMTN-135}.

\section{Image processing}
\label{section-image-processing}

\begin{figure}[h]
\includegraphics[width=\textwidth]{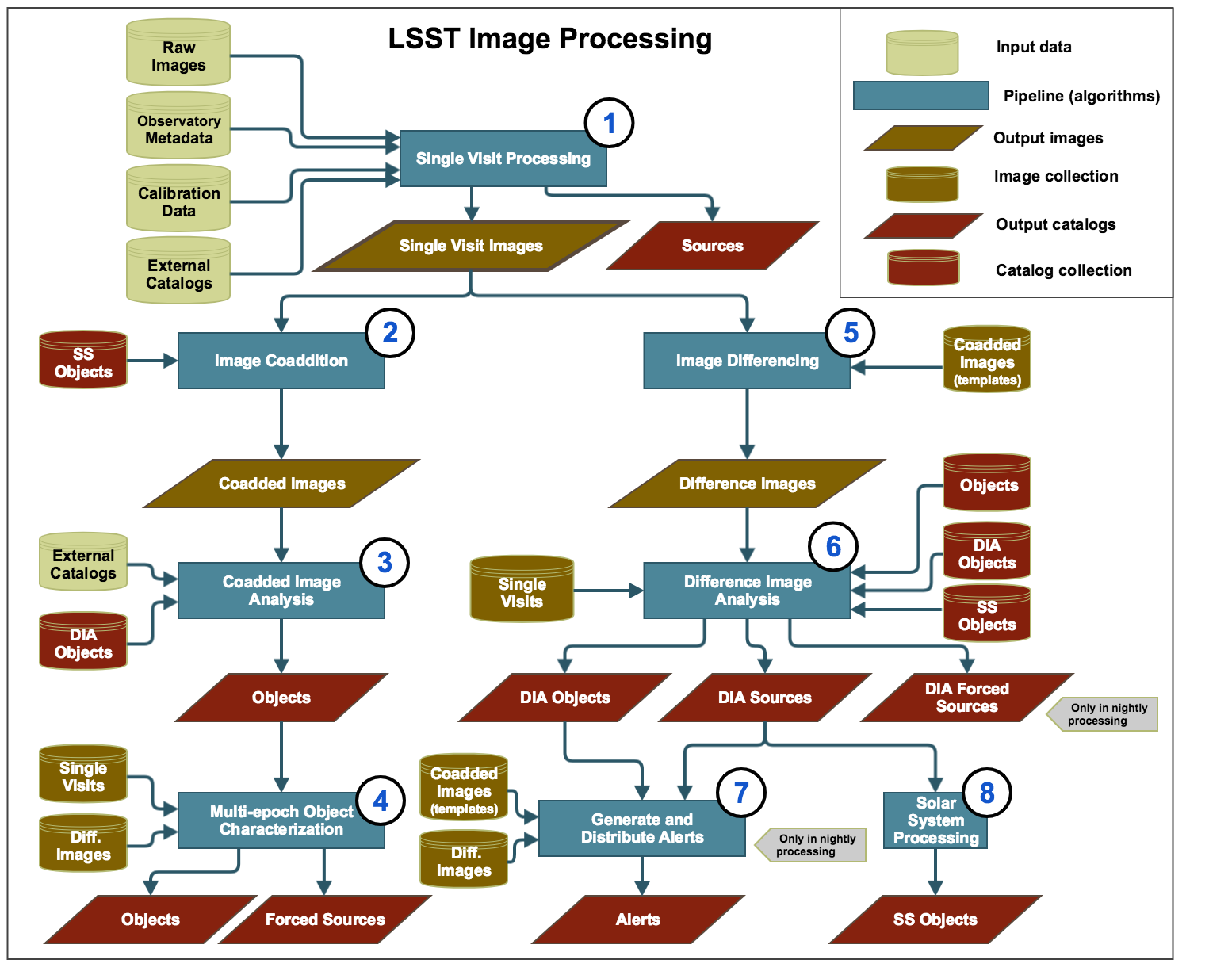}
\caption{Illustration of the conceptual design of the LSST Science Pipelines for image processing. For details see\,\cite{LSE-163}.}
\label{fig:science-pipelines}
\end{figure}

Detection of astrophysical objects and measurement of their properties from the input images is implemented in the form of a \textit{processing pipeline}, a set of connected processing elements wherein each element performs a specific analysis task. Pipeline tasks receive their inputs from either the outputs of upstream tasks or external data sources such as the input raw data or reference catalogs (see Figure \ref{fig:science-pipelines}). The result of this production is one or more data sets which are either intermediate products consumed as inputs to downstream tasks or final products included in the annual Data Release\,\cite{bosch-pipelines}.

The Rubin LSST Science Pipelines are composed of about 80 different kinds of tasks at the time of this writing, which are all implemented on top of a common algorithmic code base and purpose-built middleware components. The Data Butler (hereafter the "Butler") is the software system that abstracts the data access details (including data location, data format and access protocols) from the pipeline developers and users of the released data, and comprises one of the components of the complete processing pipeline\,\cite{2022SPIE12189E..11J}.

An instance of a Data Butler is called a \textit{repository}, which is composed of a \textit{registry} and a \textit{datastore}. Datasets are organized in the registry according to astronomical concepts such as exposure, detector, or band, without knowledge of how those data sets are persisted or where they are located. In-memory Python objects are serialized by the Butler datastore abstraction and written to a storage system. The Butler datastore also provides functionality for reading these objects from storage, delivering a recreated Python object to the calling code. Pipeline tasks use the Butler's Python API to retrieve specific data sets needed as their inputs as well as to orchestrate storage of their outputs. The Butler registry is implemented on top of a relational database (currently both PostgreSQL\,\cite{postgresql} and SQLite\,\cite{sqlite} are supported) and the datastore lays on top of a physical storage system which exposes one of the currently supported access protocols, namely POSIX, S3\,\cite{s3}, Google Cloud Storage\,\cite{GCS} and webdDAV\,\cite{webdav}.

A directed-acyclic graph known as a \textit{QuantumGraph} is generated by an algorithm which takes as inputs a file describing the tasks that compose a particular pipeline, the configuration of each task, the dependencies between them and the criteria to select the input data sets on which the pipeline is to be applied.

To execute a pipeline, the associated QuantumGraph, typically composed of hundreds of thousands of nodes, is first translated into a format expected by a workflow management system. The Batch Production System\,\cite{bps} performs this translation and can target various workflow execution systems such as PanDA\,\cite{panda,Maeno_2014}, HTCondor\,\cite{htcondor} and Parsl\,\cite{babuji19parsl}. The selected workflow management system interacts with a batch system which orchestrates the execution of the pipeline in the proper sequence. This additional level of abstraction allows the Rubin Observatory to perform its image processing campaigns using various mature workflow and workload management systems.

\section{Processing infrastructure}
\label{section-processing-infrastructure}

The image processing pipelines must be executed on a computing and storage infrastructure that allows for execution at the required scale for production of a Data Release. The compute capacity required to perform the annual reprocessing campaigns is estimated to be 10,000 CPU cores in the first year of operations and is expected to increase by one order of magnitude by the end of the LSST survey ten years later\,\cite{DMTN-135}.

In this section we present the deployment structure of the Observatory's distributed data processing infrastructure.

\subsection{Data Facilities}
\label{subsection-data-facilities}

\begin{figure}[h]
\includegraphics[width=\textwidth]{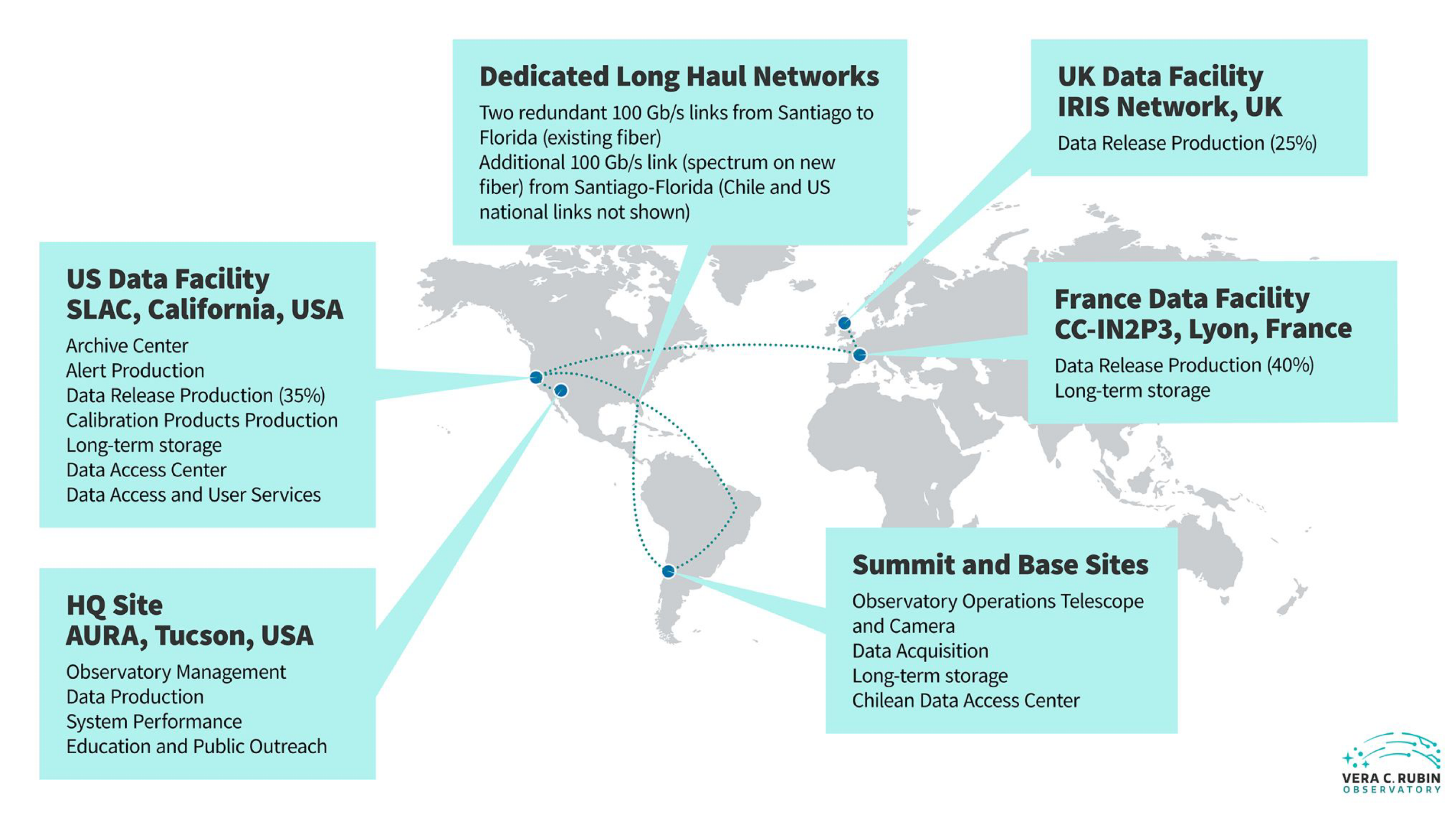}
\caption{Images flow from the Summit Site, where the telescope is located in Chile, to the Base Site and then to the three Rubin Data Facilities which collectively provide the computational capacity for processing the images taken by the Observatory for the duration of the survey.}
\label{fig:data-facilities}
\end{figure}

The LSST image processing pipelines will be executed at three facilities, with one located in the USA and two in Europe (see Figure \ref{fig:data-facilities}).

SLAC National Accelerator Laboratory\footnote{https://www.slac.stanford.edu} is the home of the Rubin Observatory data archive center and hosts the US Data Facility (USDF). The LSST:UK Science Centre\footnote{https://www.lsst.ac.uk} hosts the UK Data Facility (UKDF) and IN2P3 / CNRS Computing Centre (CC-IN2P3)\footnote{https://cc.in2p3.fr} the France Data Facility (FrDF). The USDF is the only facility to perform Prompt Processing and contributes 35\% of the computing resources required to perform Data Release Processing, with the UKDF contributing 25\% and the FrDF the remaining 40\%.

Multiple copies of the raw data will be stored among these facilities. The USDF will permanently store a complete copy of both the raw data as well as data products resulting from the annual processing and will be the primary institution responsible for dissemination of data products to LSST Science Collaborations.\footnote{Released data products will also be served by independent data access centers distributed around the world.} Other copies of the data will be spread among the Data Facilities in order to mitigate the effect of site-specific data loss\,\cite{RTN-054}.

Each Data Facility operates several services for the needs of data processing, including a Butler repository, a compute farm controlled by a workload management system for execution of the pipelines as well as some supporting services. Importantly, the Butler repository at each facility only contains records of the  data sets located at the facility's storage system and, for efficiency, tasks executed at each facility only process data sets located onsite or at a short network distance away\,\cite{DMTN-189}.

\subsection{Software distribution}
\label{subsection-software-distribution}

The term \textit{LSST Science Pipelines}\footnote{https://pipelines.lsst.io} designates a coherent software distribution which includes the set of packages developed by the Rubin Observatory's Data Management team, as well as by third-parties for processing the LSST data. The distribution includes image processing algorithms to perform, among other functions, single-frame processing, calibration, image coaddition, processing of coadded images, difference image analysis as well as production of catalog data.

Stable and weekly releases of the LSST Science Pipelines are made available through a CernVM-FS based software distribution network\footnote{https://sw.lsst.eu}\,\cite{cvmfs,Blomer_2012} in the form a conda-based set of packages\,\cite{conda}, as well as executable container images including OCI-compatible (e.g. Docker\,\cite{docker}) and Apptainer images\,\cite{apptainer}.

The three Data Facilities' compute farms are configured to mount CVMFS as a read-only file system on all of their compute nodes. This helps with reproducibility by ensuring that all the compute nodes used for the annual processing campaigns, whatever their location, execute a bit-by-bit identical copy of a given release of the pipeline software. For official processing campaigns, Linux is the agreed upon operating system for execution of the LSST Science Pipelines at all the Data Facilities.\footnote{The LSST Science Pipelines also run on macOS.}

\subsection{Data distribution}
\label{subsection-data-distribution}

Prior to the annual processing campaigns, raw images are replicated from the USDF to the European facilities, stored into their local storage systems and registered into the facility's local Butler repository.\footnote{This replication is performed shortly after images taken by the Observatory arrive at the USDF for Prompt Processing.} Rubin utilizes Rucio\,\cite{rucio2019} for driving this continuous data replication. Rucio is a policy-based data management system which orchestrates the replication of datasets to satisfy configured rules (e.g. one copy of this particular kind of data is required at each facility), submitting file transfer requests to an FTS3\,\cite{FTS3} service which manages the execution of those transfers. Transfer requests typically require that the storage systems at the destination facility download the files from the source facility over a confidential communication channel. Inter-facility file transfers for Rubin data use secure HTTP as transport protocol and are performed by agents authenticated via X.509 certificates. Rubin's Rucio and FTS3 instances are located at the US Data Facility.

Upon the success of data replication, automated actions are triggered to ingest the replicated datasets into the destination facility's local Butler repository. Middleware developed by the Rubin team runs at the USDF listening to data replication events emitted by Rucio. The middleware then emits messages which trigger ingestion of data into the local Butler repository at each destination Data Facility\,\cite{DMTN-213}. Final data products generated by the annual reprocessing campaigns at each processing facility are replicated to the archive site at USDF using the same mechanism and are ingested into its local Butler repository.

High-performance network links interconnect the Rubin Data Facilities. Those links are provided by ESnet\footnote{https://www.es.net} within the USA and to cross the Atlantic, and by GÉANT\footnote{https://www.geant.net}, RENATER\footnote{https://www.renater.fr} and the Janet network\footnote{https://www.jisc.ac.uk/janet} within Europe.

\subsection{Distributed processing campaigns}
\label{subsection-distributed-processing-campaigns}

Rubin will conduct annual reprocessing campaigns to produce the Data Releases, exploiting the computing resources available at the Data Facilities according to the agreed workload share described in section \ref{subsection-data-facilities}.

A QuantumGraph specific to each facility is centrally generated according to the input data residing there. The workflow associated with that graph is sent to the PanDA system which orchestrates the execution of the batch jobs by preparing and submitting pilot jobs to each Data Facility's compute element. A Data Facility compute element is composed of a workload management system (typically SLURM\,\cite{slurm}) and a gateway service (typically Nordugrid's ARC\,\cite{arc}) which securely exposes the facility's batch farm for PanDA direct usage. Each batch job is configured such that it can find the local Butler repository, utilize the appropriate release of the LSST Science Pipelines as well as the specific configuration of the pipeline tasks it must execute.

The USDF operates a dedicated PanDA instance which orchestrates job execution at all Rubin Data Facilities\,\cite{panda-rubin}. Each Data Facility's compute gateway receives PanDA jobs and submits them for execution to the local batch farm. PanDA interacts with the gateway to manage the lifetime of each job and performs retries when required. A particular node in the QuantumGraph is tasked with the transfer of the generated data products to the local Butler repository. The execution logs generated by each job at all the Data Facilities are sent to a central service which runs at the USDF so that the Rubin production team may monitor the progress of the whole campaign.

\section{Summary}
\label{summary}
We presented a high-level view of the systems the Rubin Observatory has deployed for processing sky image data in three data facilities on two continents over the course of the ten year-long survey. Key components of those systems were also briefly presented, the interactions between them and select external tools developed for other science projects that Rubin adopted for tackling the data processing challenges it faces.

\section{Acknowledgments}

This material is based upon work supported in part by the National Science Foundation through Cooperative Agreement AST-1258333 and Cooperative Support Agreement AST-1202910 managed by the Association of Universities for Research in Astronomy (AURA), and the Department of Energy under Contract No. DE-AC02-76SF00515 with the SLAC National Accelerator Laboratory managed by Stanford University. Additional Rubin Observatory funding comes from private donations, grants to universities, and in-kind support from LSSTC Institutional Members.

\bibliography{references}

\end{document}